\documentclass[number,sort&compress,times]{{elsarticle}}
\usepackage[utf8]{inputenc}
\usepackage{amssymb}
\usepackage{amsthm}
\usepackage{amsmath}
\usepackage{graphicx}
\usepackage{multirow}
\graphicspath{ {./} {./images/} }
\usepackage[skip=0.5\baselineskip]{caption}
\usepackage{float}
\usepackage{times}
\usepackage{xcolor}
\usepackage{natbib}
\usepackage[%
        colorlinks=true,urlcolor=blue,citecolor=blue,
        pdfpagelabels=true,hypertexnames=true,
        plainpages=false,naturalnames=true
        ]{hyperref}

\usepackage{newfloat}
\DeclareFloatingEnvironment[name={Table}]{mytable}

\usepackage[version=4]{mhchem}
\newcommand{\red}[1]{\textcolor{black}{#1}}

\begin{document}

\title{Computational Design of Corrosion-resistant and Wear-resistant Titanium Alloys for \red{Orthopedic Implants}}
\author[1]{Noel Siony}
\author[2]{Long Vuong}
\author[2]{Otgonsuren Lundaajamts}
\author[3]{Sara Kadkhodaei\corref{cor1}}
\ead{sarakad@uic.edu}

\cortext[cor1]{Corresponding author}
\address[1]{Department of Physics,
University of Illinois Chicago, 845 W. Taylor St., Chicago, IL 60607 USA}
\address[2]{Mechanical \& Industrial Engineering,
University of Illinois Chicago, 842 W. Taylor St., Chicago, IL 60607 USA}
\address[3]{Civil, Materials, and Environmental Engineering,
University of Illinois Chicago, 2095 Engineering Research Facility,
842 W. Taylor St., Chicago, IL 60607 USA}

\begin{abstract}
Titanium alloys are promising candidates for orthopedic implants due to their mechanical resilience and biocompatibility. Current titanium alloys in orthopedic implants still suffer from low wear and corrosion resistance. Here, \red{we present a computational method for optimizing the composition of titanium alloys for} enhanced corrosion and wear resistance without compromising on other aspects such as phase stability, biocompatibility, and strength. We use the cohesive energy, oxide formation energy, surface work function, and the elastic shear modulus of pure elements as proxy descriptors to guide us towards alloys with enhanced wear and corrosion resistance. For the best-selected candidates, we \red{then} use the CALPHAD approach, \red{as} implemented in the Thermo-Calc software, to calculate the phase diagram, yield strength, hardness, Pourbaix diagram, and the Pilling-Bedworth (PB) ratio. These calculations are used to assess the thermodynamic stability, biocompatibility, corrosion resistance, and wear resistance of the selected alloys. Additionally, we provide insights about the role of silicon on improving the corrosion and wear resistance of alloys. 
\end{abstract}
\maketitle
\section{Introduction}\label{sec:introduction}

Orthopedic implants have significantly improved the quality of life for people who have irreversibly injured their bones \red{\cite{Lee2008,Marsh2021,NGUYEN2015,Lehil2014}}. In the past, a significant injury to the hip or knee was recovered by amputating the affected limbs~\cite{Hernigou2013-kh}. Initial orthopedic implants were very different from what we use today. More akin to metal casts, early implants used plates and screws to reinforce damaged bones rather than completely replace the damaged bone with a new artificial part. It was not until later that science had progressed far enough to make the concept of a complete replacement possible for patients. The first knee implants utilized ivory as the primary construction material \red{\cite{Papas2018}}. The success of \red{the material for} an implant depends on several factors: biocompatibility with the body, the strength of the material, ability to resist wear, and the ability to resist the corrosive conditions of the body. At the time, ivory was the best material that fit these criteria as it can withstand compression forces quite well and is naturally resistant to corrosion inside the body \cite{Szostakowski2017}. Additionally, studies done on patients long after they had been given an ivory implant showed that the bone fused well with the ivory implant in most areas \cite{Szostakowski2017}. 

Later, ceramic implants were tested because they provided higher wear resistance. Today, ceramic implants are still used for patients who reject metal implants due to immune reactions. However, the major drawbacks to ceramic implants come from their brittle nature and inability to take repetitive impacts \red{(dynamic loading)} from activities such as jogging \cite{Wright2020}. Ceramic implants bring a higher chance of fracture and, in the worst cases, the complete shattering of the implant. Today, a commonly used ceramic is alumina due to its great ability to withstand compression forces and wear. However, alumina suffers from a weakness to tensile stresses like other ceramics do \cite{Piconi2003-jt}. Zirconia is a common alternative to alumina as it shares hardness properties comparable to alumina while having incredible crack resistance. A composite material was developed composed of zirconia introduced into an alumina matrix called zirconia toughened alumina or ZTA \cite{Gadow2010}. This composite combines the best features of its base materials and offers a material that is highly resistant to wear and cracking while maintaining its high toughness \cite{Merola2019}. Though ZTA excels in these categories, it is prone to aging in the presence of water; a form of deterioration that leads to lowered strength and a higher risk of fracture \cite{gutknecht2007}.

Metal implants became available as the metal industry evolved and more manufacturing processes became available and refined \red{\cite{TADDEI2004683,Majumdar2018,Izmir2019,RAZA2022649,SINGH2022612,NAG2005357}}. Early implementations of metal implants suffered from a lack of quality control leading to low wear resistance and fracturing \cite{Hermawan11}. The low wear resistance lead to more significant implant deterioration and cause free-floating metal ions to enter the bloodstream. For patients with weak kidneys, this heightened metal level in the bloodstream could result in metal poisoning as they cannot filter out the metal fast enough \cite{Hermawan11}. 
\red{To avoid metal poisoning, metal implants with higher wear resistance are needed. Aside from high wear resistance, the desired characteristics are high strength, low modulus, and excellent corrosion resistance. The combination of low elastic modulus and high strength of the implant results in a more uniform distribution of stresses between the bone and the implant \cite{Heary2017,Merola2019}. }

Typical metals for orthopedic implants are cobalt-chrome alloys (CoCr), stainless steel, and titanium alloys. For a long time, CoCr was one of the primary materials used in the construction of surgical and dental implants due to its ability to resist corrosion better than stainless steel and ceramics. CoCr utilizes its oxide layer, \ce{Cr2O3}, to protect itself from corrosion. Its high corrosion resistance combined with wear resistance, biocompatibility, and strength, makes CoCr the alloy of choice for implants \cite{Shah2018-pm}. Even though CoCr is one of the most wear-resistant and corrosive-resistant alloys for implants, if the oxide layer does get worn down and cobalt and chromium particles are exposed to the patient’s bloodstream, it will result in severe immune reactions. While the effects of low amounts of Co and Cr particles in the body are still mostly unknown, high amounts of these elements are highly toxic to humans \red{\cite{Delaunay2010-kh,Dalal2012}}. Surgical grade stainless steel (316L) is mainly used for non-permanent applications such as stabilizing broken bones to help the body heal. In some cases, these implants are replaced with a CoCr or titanium-alloy-based implant if the situation demands it. Stainless steel transient usage is because stainless steel does not resist corrosion and is not as strong as titanium alloys \cite{Merola2019}. Alloying stainless steel with other elements does improve corrosion resistance but it is still outclassed by many other materials. 

Titanium alloys have excellent material properties and high corrosion resistance~\cite{WANG1996,Guillemot2005,KAUR2019844}. Like CoCr, titanium naturally forms an oxide when inside the human body that protects it from corrosion resistance \cite{Textor2001}. Grade four commercially pure titanium (CP-Ti) and Ti-6Al-4V (Ti-64) are the two most commonly used forms of titanium for medical implants \red{\cite{Elias2008,WANG1996}}. Titanium’s success as an implant comes from its superiority in corrosion resistance, biocompatibility, and high strength to weight ratio\red{~\cite{Elias2008,Sidambe2014,Kirmanidou2016,Ragone2020,Dalal2012,Hanawa,ABDELHADYGEPREEL2013407}}. However, titanium suffers from lower wear resistance than CoCr and more often results in wear debris in patients \cite{Merola2019}.
The debris poses a considerable risk to patients since vanadium is highly cytotoxic \cite{Li2010-gj}. Many substitutions to vanadium have been made to create a safer alloy. For example, Ti-6Al-7Nb substituted out the cytotoxic vanadium for the much safer alternative of niobium which had the added benefit of increasing corrosion resistance \cite{Li2010-gj}. Ti-5Al-2.5Fe is another alloy created for a similar purpose. Ti-5Al-2.5Fe is an improvement on Ti-6Al-4V in almost all fields \cite{SIMSEK2019357}. Surface modification of titanium alloys have also been reported to improve the corrosion resistance, such as the use of \ce{SiO2} oxide for coating the Ti6Al7Nb alloy and the CP-Ti titanium (Grade 4)~\cite{Basiaga2014,BIENIAS2009}. However, even with all these improvements, all Ti-alloys still suffer from the same problem that all implants suffer from: the constant friction forces generate considerable amounts of metal debris inside the body over a long time \cite{SIMSEK2019357,FAIS2012}, and new modifications of the alloy composition to improve wear resistance is desirable.

\red{This report aims to present a computational method for optimizing the composition of titanium alloys for increased wear and corrosion resistance. The presented method and alloy optimization results serve a two-fold purpose: 1) To provide a high-throughput method for down-selecting alloys with improved corrosion and wear resistance, which can be utilized to guide the experimental design of orthopedic implants. 2) To deepen our understanding of the effect of different alloying elements on increasing the mechanical compatibility, wear resistance, and corrosion resistance of alloys.}
The presented computational approach is used for assessing the stability, mechanical biocompatibility, wear resistance, and corrosion resistance of several titanium alloys. We adopt a two-tier approach: In the first tier, we use fundamental atomic and electronic attributes for a high-throughput selection of alloying elements that increase wear and corrosion resistance. In the second tier, we use the CALPHAD databases and tools integrated within the ThermoCalc software to predict the thermodynamic stability, corrosion resistance, wear resistance, and mechanical compatibility of the selected ternary titanium alloys. We show the improvement of corrosion resistance among the selected alloy based on their reduced Pilling-Bedworth (PB) ratio, defined as the volume of formed oxide phase(s) divided by the volume of alloy phase(s) that formed the oxides. Enhanced wear resistance is assessed based on the increased hardness of the selected alloys compared to the benchmark system of Ti-6Al-4V (mole\%). 

In section \ref{sec:method}, we elaborate on the methodology used in this work. In section \ref{sec:results}, we utilize the methodology to select ternary titanium alloys with improved wear and corrosion resistance. We study the stability, biocompatibility, wear resistance, and corrosion resistance of the selected alloys using the CALPHAD approach. In section \ref{sec:silicon}, we study the role of increased silicon concentration in improving the wear and corrosion resistance of alloys and suggest a pathway toward the use of silicon alloys for medical implants. \red{In section \ref{sec:discussion}, we discuss the limitations of the presented method compared to available studies for the dynamic corrosion behavior and tribocorrosion phenomena in alloys for orthopedic implants. Additionally, we compare the findings of this work with other experimental and clinical studies of the mechanical properties, wear resistance, and corrosion resistance of Ti-alloys for implants.} Finally, in section~\ref{sec:conclusion}, we present conclusive remarks about the effect of different alloying elements on improving the hardness and corrosion resistance of the alloys in our study.  

\section{Method}\label{sec:method}
We use a two-tier process to design Ti-alloys with enhanced wear and corrosion resistance. The complexity and accuracy of our analysis increase progressively, moving from tier one to two. In tier one, a rapid screening that covers a wide range of alloying elements is performed based on simple fundamental parameters as descriptor proxies for corrosion and wear resistance. To this end, we use available density functional theory (DFT) data for the relevant fundamental parameters, including the cohesive energy, oxide formation energy, surface work function, and elastic shear modulus. A similar approach based on cohesive energy and oxide formation \red{energy} was initially proposed by Markus~\cite{MARCUS19942155} but has only been examined for pure metals. Taylor and coworkers suggested that an integrative approach that can benefit from these descriptor proxies should be further developed ~\cite{Taylor2018}. Here, we extend Markus’ approach to alloy systems and add the surface work function as an additional descriptor. Our approach provides new insights into the suggested integrated computational materials engineering (ICME) approach by Taylor and co-workers. 

In tier two, the corrosion and wear resistance of selected alloys from tier one will be assessed based on comprehensive thermodynamic data and sophisticated optimization techniques employed in the Thermo-Calc Software TCTI3 Ti-alloys database \red{\cite{ANDERSSON2002273}}. We use the Equilibrium Calculator for phase diagram and Pourbaix diagram calculations and the Property Model Calculator for yield strength and hardness calculations. 

\section{Results}\label{sec:results}
\subsection{High-throughput Screening of Ti alloys}\label{sec:dft}
We use three different DFT parameters, calculated for transition metals and metalloids elements, as proxy descriptors to optimize the alloys for enhanced corrosion resistance: 1) the cohesive energy, which is the measure of the metal-metal (M-M) bond strength, as a proxy descriptor of dissolution resistance, 2) the oxide formation energy, which is the measure of the metal-oxygen (M-O) bond strength, as a proxy descriptor for oxide scale formation tendency, and 3) the surface work function, which is the electrostatic work needed to transport the charged electron through the dipole layer of the metal surface undergoing oxidation. A large surface work function shows the electrochemical nobility of the metal and favors the galvanic corrosion resistance of the metal surface. The DFT parameters are collected from the Materials Project database~\cite{Jain2013} and are provided in Supplemental Table 1. The oxidation formation energy is the formation energy per atom for the most stable metal oxide available in the Materials Project database. The surface work function is the weighted average over different surface orientations. The cohesive energy is the difference between the bulk energy and the sum of total DFT energy of isolated atoms in the bulk, obtained from the Materials Project database.

For increasing the corrosion resistance, our strategy is to select metal components that combine the oxide scale formation tendency, dissolution resistance, and galvanic corrosion resistance based on the aforementioned proxy descriptors. In other words, we optimize the alloy composition through a synergistic operation of alloying elements as passive (or protective) scale promoters and dissolution blockers. We combine alloying elements that promote the formation of a passive oxide scale (i.e., selecting elements with high oxidation formation tendency) with those with a high dissolution resistance under the human body chemical conditions (i.e., selecting elements with high cohesive energy and work function). Thereby, we identify scale-forming alloying elements that can combine the following attributes: high oxidation tendency in the base alloy to form a stable oxide scale combined with high cohesive energy and work function to mitigate dissolution and galvanic corrosion.

\begin{figure}[!h]\includegraphics[width=\textwidth]{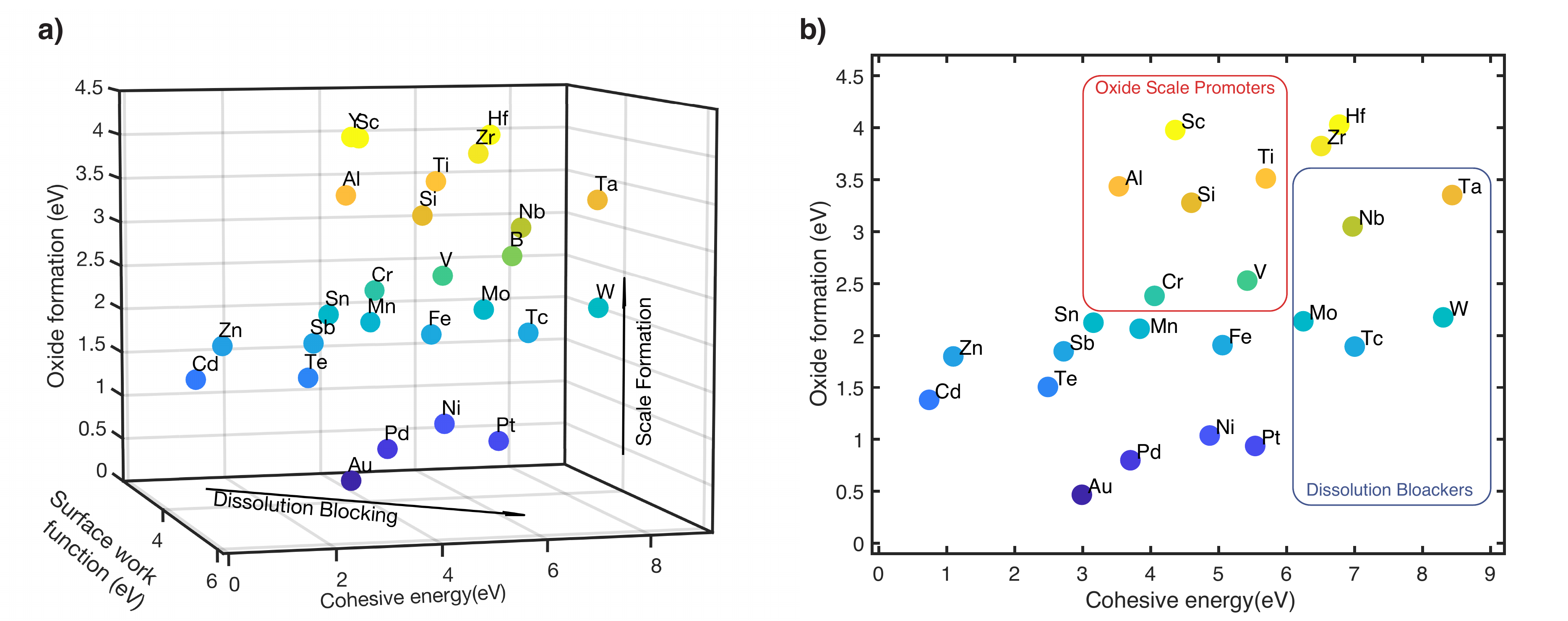}
\caption{Fundamental parameters of potential alloying elements (transition metals and metalloids). a) Oxide formation energy versus the cohesive energy and the weighted surface work function for several possible alloying elements. The arrows illustrate the increased dissolution blocking and scale formation capability of different alloying elements. b) Cohesive energy (M-M bond strength) as a proxy descriptor for metal dissolution resistance versus oxide formation energy (M-O bond strength) as a proxy for oxide scale formation tendency.
}
\label{fig:DFT1}
\end{figure}

\begin{figure}[!h]\includegraphics[width=\textwidth]{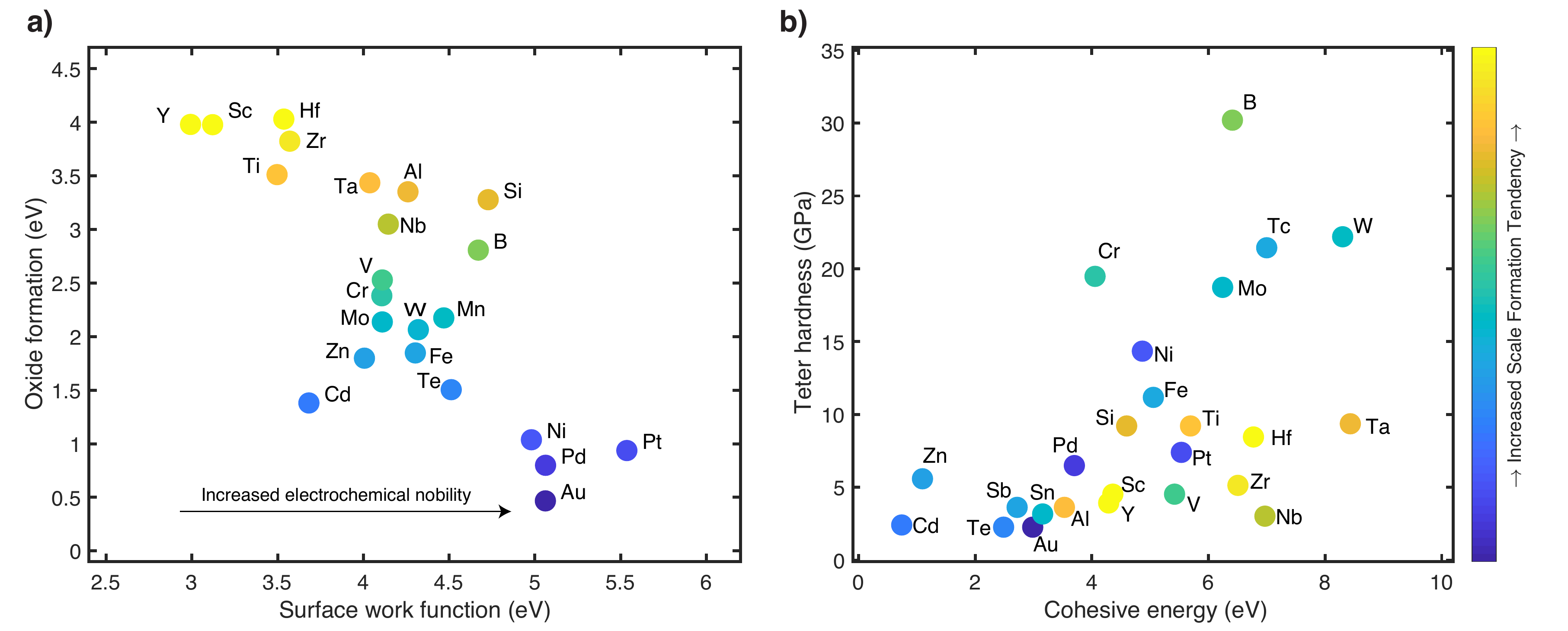}
\caption{Fundamental parameters of potential alloying elements (transition metals and metalloids). a) Oxide formation energy as a proxy descriptor for the scale formation tendency versus the weighted surface work function as a proxy descriptor for the galvanic corrosion resistance.
 b) DFT-based Teter hardness as a proxy descriptor for wear resistance versus the cohesive energy as a proxy descriptor for corrosive dissolution resistance.
}
\label{fig:DFT2}
\end{figure}

Figure~\ref{fig:DFT1} (a) shows the oxidation formation energy in terms of the cohesive energy and surface work function for several possible alloying elements. Increasing the surface work function and cohesive energy favors resistance to corrosive dissolution while increasing the oxide formation tendency favors oxide scale formation. Figure \ref{fig:DFT1} (b) illustrates the oxide formation energy against the cohesive energy. Elements with high oxide formation energies and relatively low cohesive energies are good oxide scale promoters (see Figure \ref{fig:DFT1} (b)). Classical examples are Cr and Al in superalloys that form protective \ce{Cr2O3} and \ce{Al2O3} scales~\cite{Bradford1993}. On the other hand, elements with high cohesive energies and relatively low oxide formation energies are good dissolution blockers (see Figure \ref{fig:DFT1} (b)). To combine the two adversarial effects in the alloy, we select the second alloying element from the oxide scale promoters including Cr, V, Si, Al, and Sc, and the third alloying element from the dissolution blockers including Ta, Nb, Mo, and W. We also consider Zr and Hf as the third alloying elements because they combine the adverse effects of oxide formation and dissolution blocking. Figure \ref{fig:DFT2} (a) shows the oxide formation energy against the surface work function. Among our selected alloying elements, Sc has the lowest surface work function. The surface work function increases for the elements in the order of  Sc$<$Zr$<$Hf$<$Mo$<$Cr$<$V$<$Ta$<$Nb$<$Al$<$W$<$Si. 

\begin{mytable}[!h]\includegraphics[width=\textwidth]{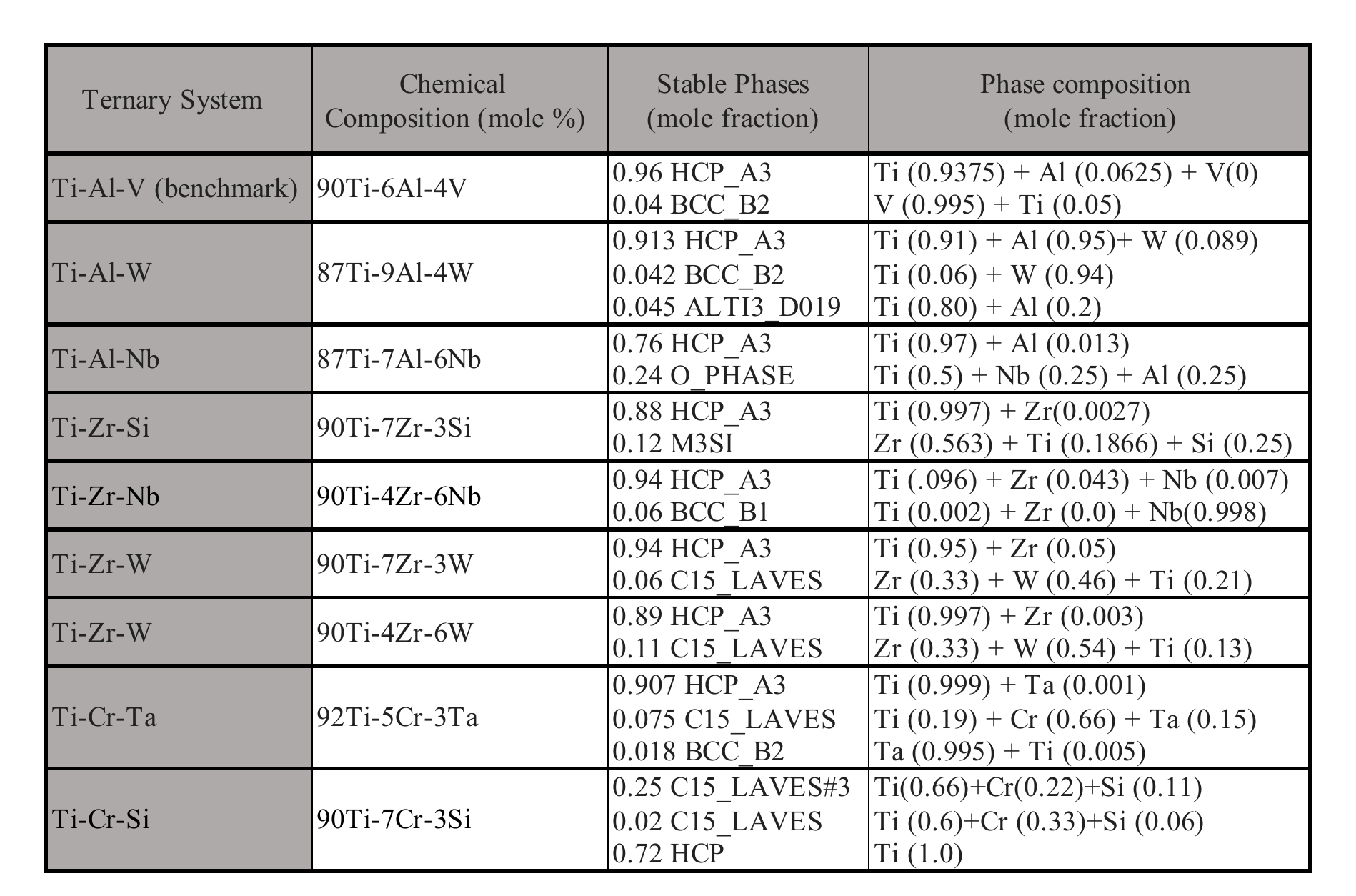}
\caption{The list of selected titanium alloys for this study. The second and third alloying elements are selected according to the fundamental atomic and electronic parameters to improve wear and corrosion resistance. The second and third columns show the selected composition of the alloys and the stable phases, respectively. The fourth column shows the composition of each stable phase.}
\label{mytable:table1}
\end{mytable}

To guide our selection towards enhanced wear-resistant alloys, we use the Teter hardness as a measure of wear-resistant \cite{teter1998}. We calculated the Teter hardness, $H$, according to $H = 0.151 G$ , where $G$ is the DFT-calculated elastic shear modulus from the Materials Project (see Supplementary Table 1). Figure~\ref{fig:DFT2} (b) illustrates the calculated Teter hardness against the cohesive energy. Among our selected alloying elements, Cr, W, and Mo have the highest hardness values. 
Table \ref{mytable:table1} shows the list of selected Ti alloys that are likely to exhibit improved corrosion and wear resistance according to the DFT parameters compared to the benchmark Ti-6Al-4V (mole \%) alloy. We provide a more detailed analysis of these alloys in the following sections. 
\subsection{Phase Diagram}\label{sec:phasediag}
\begin{figure}[!htp]\includegraphics[width=\textwidth]{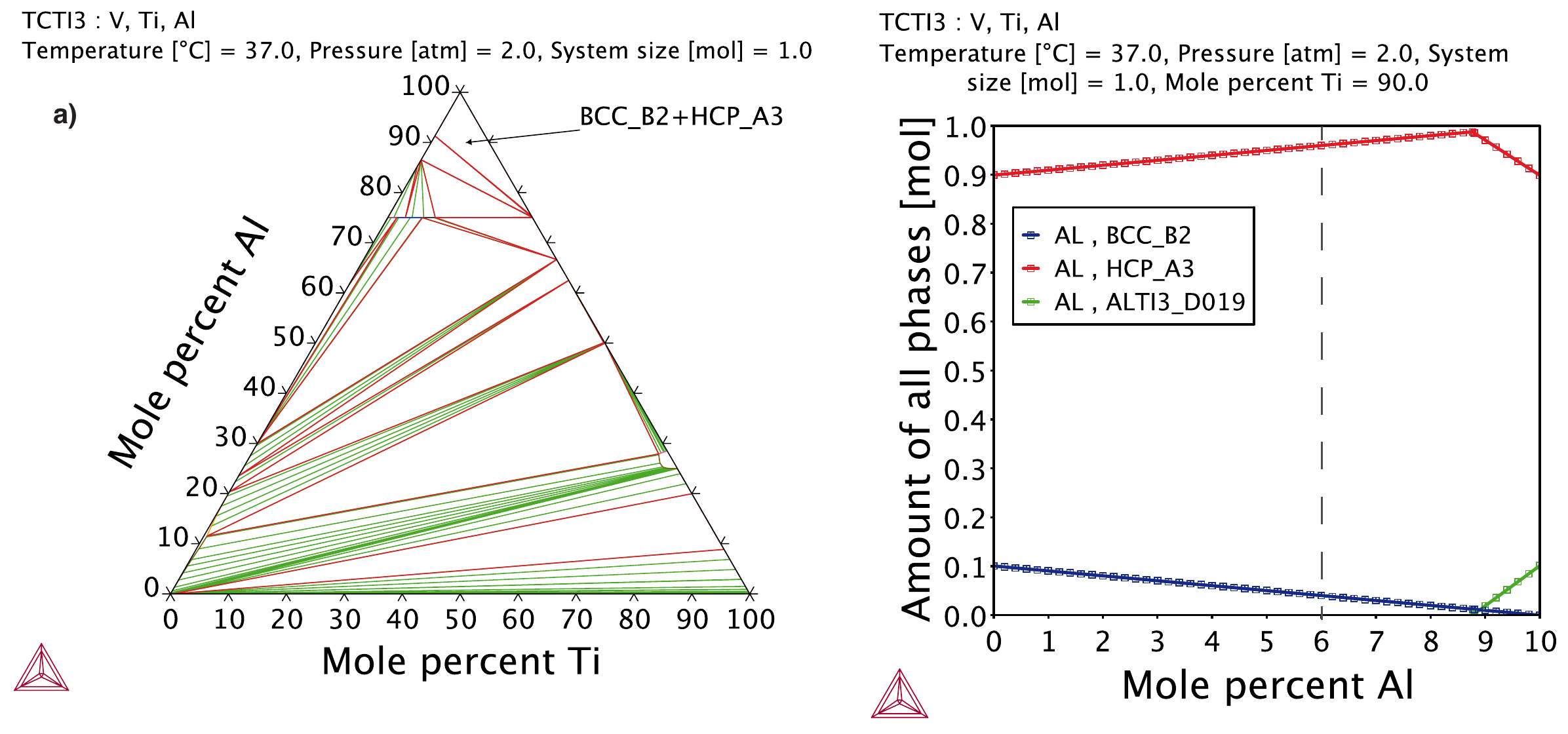}
\caption{Thermodynamic stability assessment of the Ti-Al-V ternary alloy. a) The ternary phase diagram at 37$^{\circ}$C and 2 atm, with tie lines shown by green and phase boundaries shown by red. The Ti-6Al-4V (mole \%) composition lies on the two phase region of HCP\_A3 and BCC\_B2 phases. b) The mole fraction of stable phases in terms of Al concentration for a fixed 90 mole \% Ti. The dashed line indicates the Ti-6Al-4V (mole \%) composition.}
\label{fig:TiAlVstability}
\end{figure}

\begin{figure}[!h]\includegraphics[width=\textwidth]{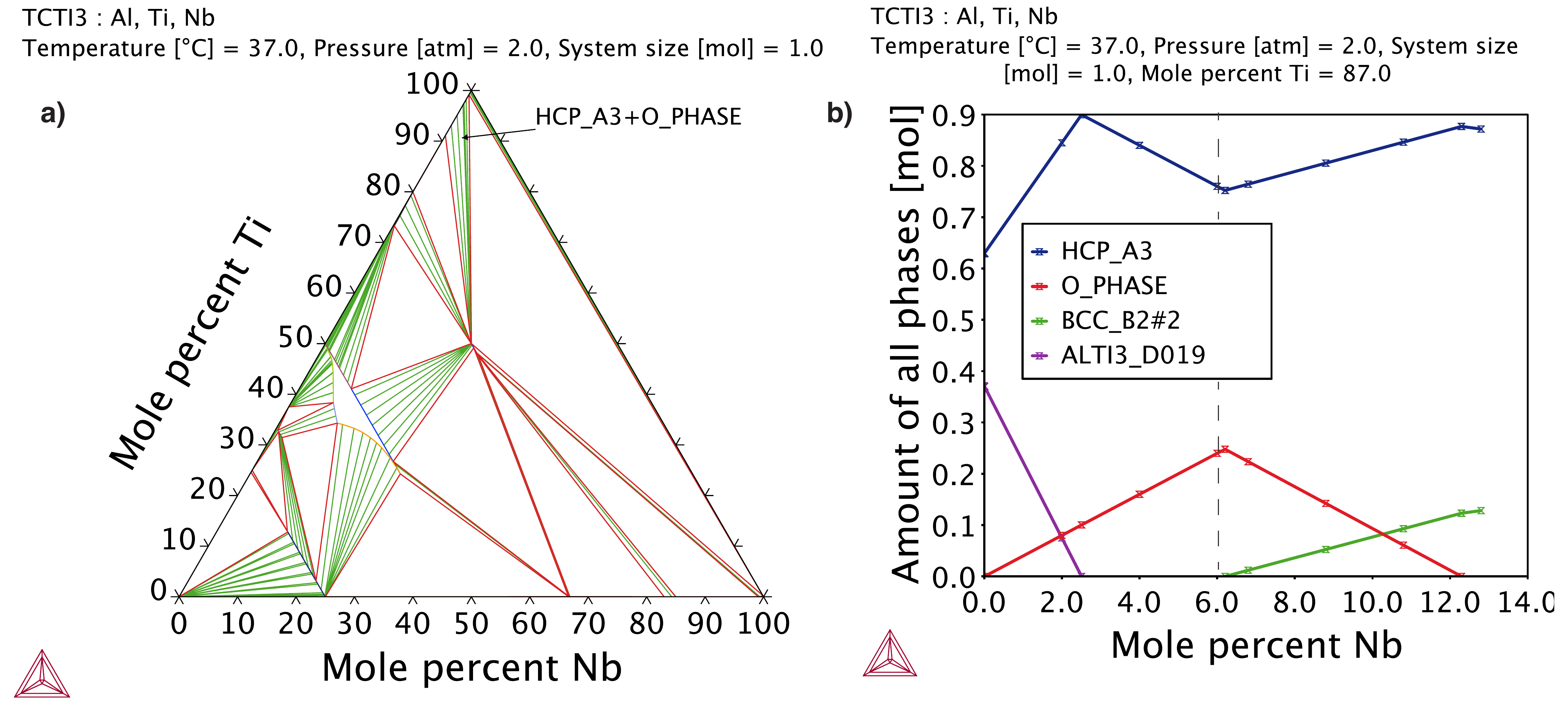}
\caption{Thermodynamic stability assessment of the Ti-Al-Nb ternary alloy. a) The ternary phase diagram at 37$^{\circ}$C and 2 atm, with tie lines shown by green and phase boundaries shown by red. The Ti-7Al-6Nb (mole \%) composition lies on the two phase region of HCP\_A3 and O (ordered orthorhombic) phases. b) The mole fraction of stable phases in terms of Nb concentration for a fixed 87 mole \% Ti. The dashed line indicates the Ti-7Al-6Nb (mole \%) composition.}
\label{fig:TiAlNbstability}
\end{figure}

\begin{figure}[!h]\includegraphics[width=\textwidth]{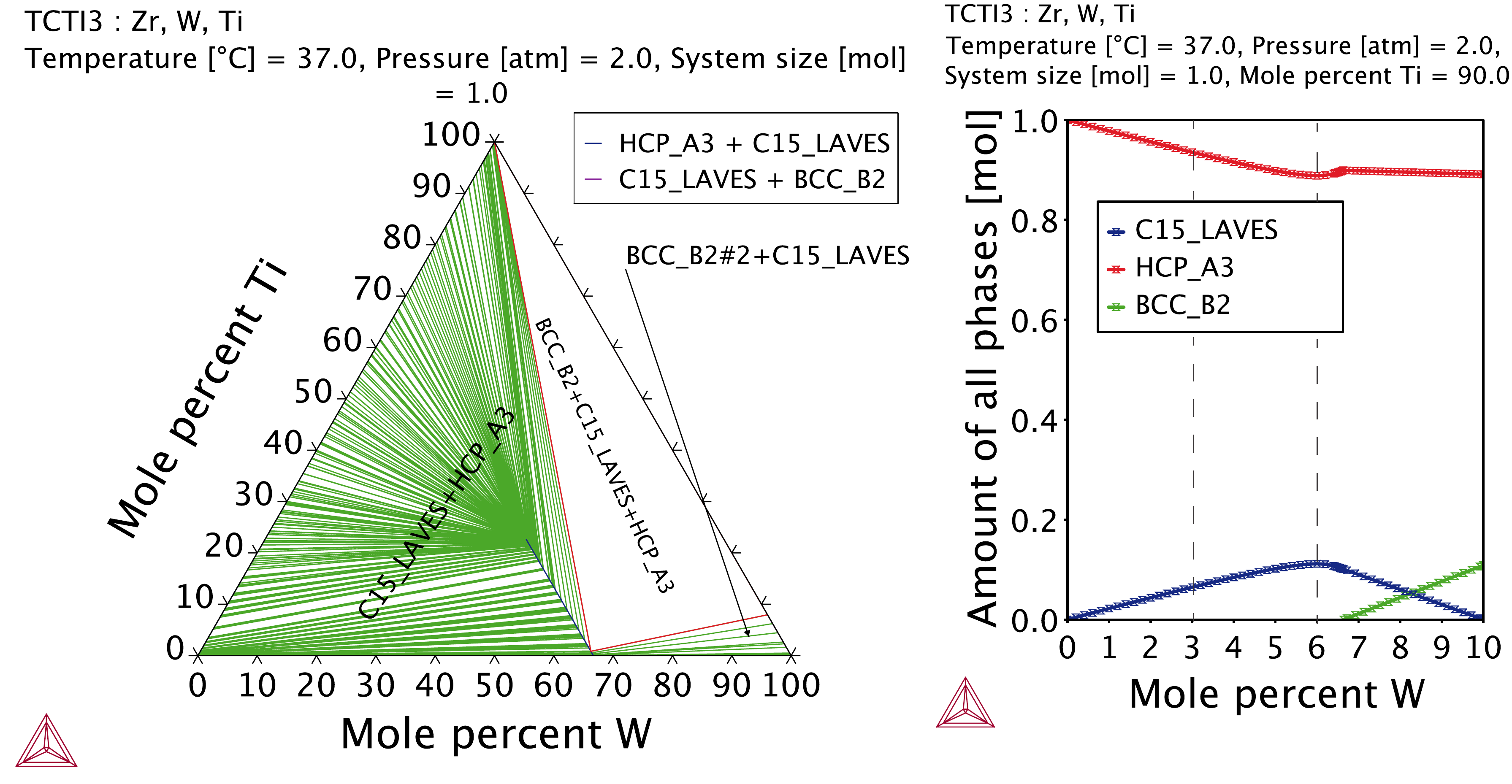}
\caption{Thermodynamic stability assessment of the Ti-Zr-W ternary alloy. a) The ternary phase diagram at 37$^{\circ}$C and 2 atm, with tie lines shown by green and phase boundaries shown by red. The Ti-7Zr-3W (mole \%) composition lies on the two phase region of HCP\_A3 and LAVES\_C15 phases. b) The mole fraction of stable phases in terms of W concentration for a fixed 90 mole \% Ti. The dashed lines indicate the Ti-7Zr-3W (mole \%) and Ti-4Zr-6W (mole \%)  compositions.}
\label{fig:TiZrWstability}
\end{figure}

We utilize the ThermoCalc software to calculate the ternary phase diagram of the selected alloys. We set the temperature to the body temperature of 37$^{\circ}$C and a pressure of 2 atm to replicate the \red{static load} in the skeletal system, specifically a typical femur bones~\cite{Davis}. Figure ~\ref{fig:TiAlVstability} (a) shows the phase diagram of the Ti-Al-V ternary alloy. Figure ~\ref{fig:TiAlVstability} (b) illustrates the mole fraction of stable phases as a function of aluminum concentration (where titanium is fixed at 90 mole \%). Figure ~\ref{fig:TiAlNbstability} (a) shows the phase diagram of the Ti-Al-Nb ternary alloy. Figure ~\ref{fig:TiAlNbstability} (b) illustrates the mole fraction of stable phases as a function of niobium concentration (where Ti is fixed at 87 mole \%). Figure ~\ref{fig:TiZrWstability} (a) shows the phase diagram of the Ti-Zr-W ternary alloy. Figure ~\ref{fig:TiZrWstability} (b) illustrates the mole fraction of stable phases as a function of tungsten concentration (where titanium is fixed at 90 mole \%). The ternary phase diagrams for other alloys of Table \ref{mytable:table1} are shown in Supplemental Figures 1 to 5. The stable phases and their corresponding mole fractions for each selected composition is summarized in Table~\ref{mytable:table1}. 
\subsection{Yield Strength and Hardness}\label{sec:strength}
\begin{figure}[!h]\includegraphics[width=\textwidth]{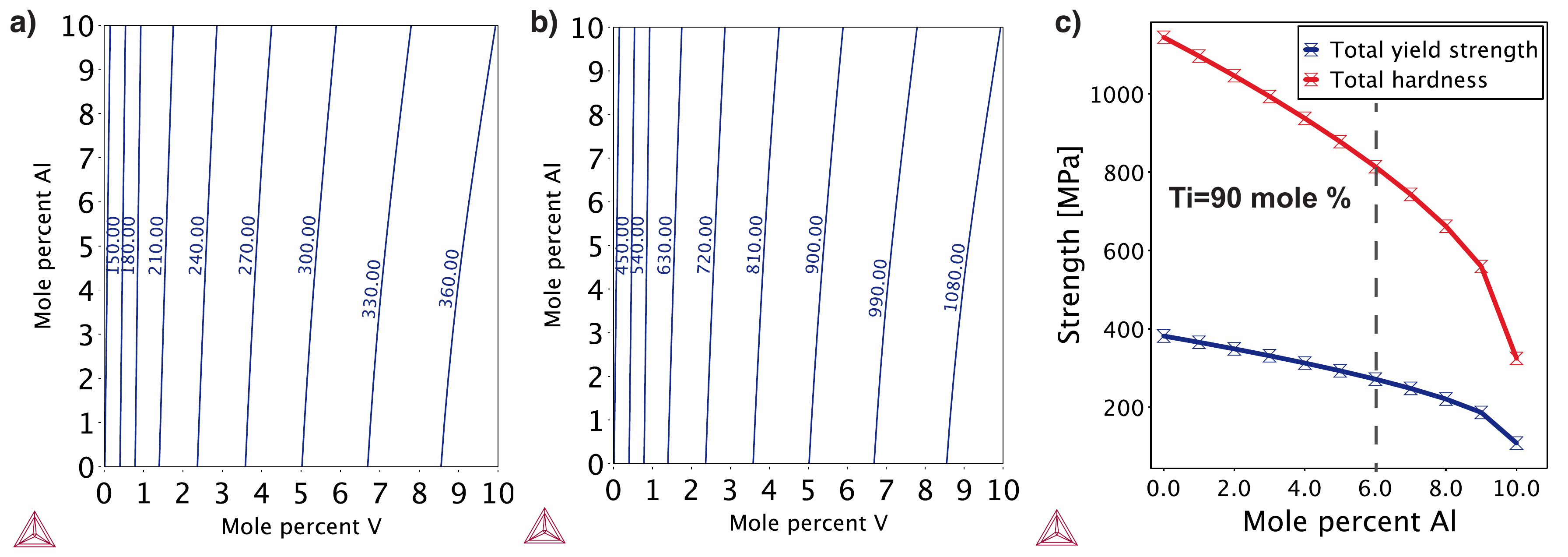}
\caption{Yield strength and hardness calculation for Ti-Al-V. a) The total yield strength and b) the total hardness contours in terms of Al and V concentrations. c) The total hardness and yield strength in terms of Al concentration for a fixed 90 mole \% Ti.}
\label{fig:TiAlVstrength}
\end{figure}
\begin{figure}[!h]\includegraphics[width=\textwidth]{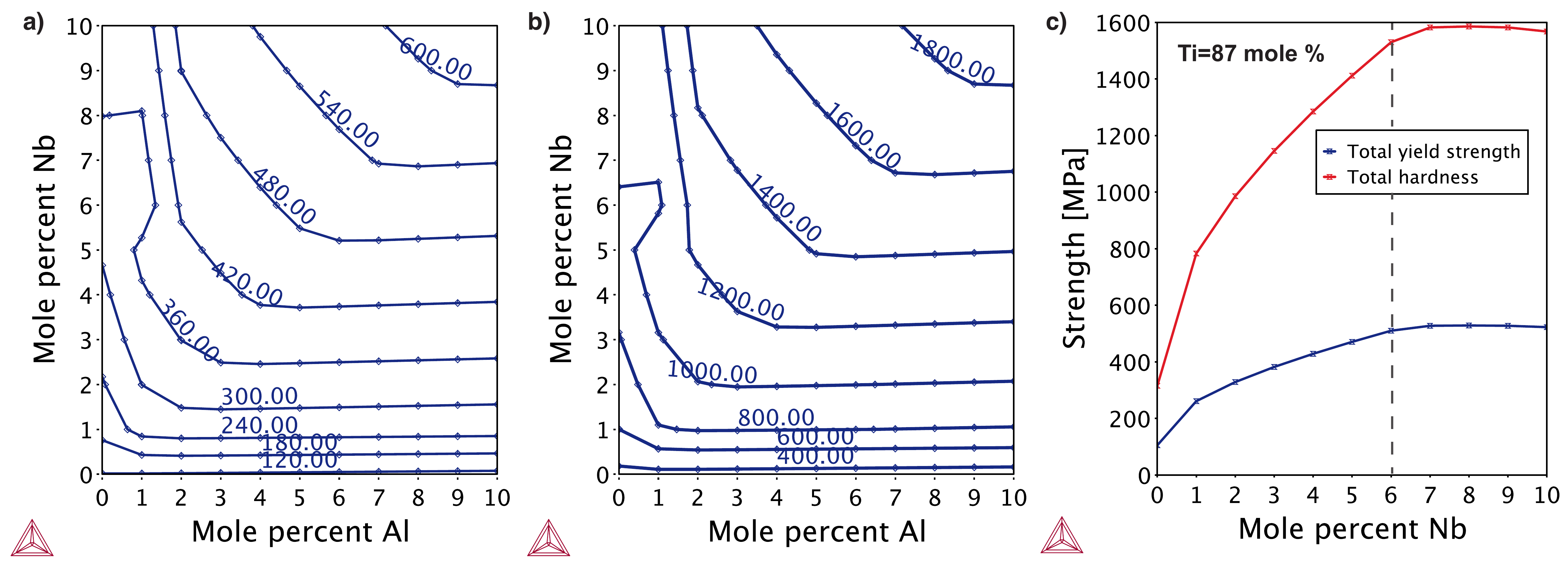}
\caption{Yield strength and hardness calculation for Ti-Al-Nb. a) The total yield strength and b) the total hardness contours in terms of Al and Nb concentrations. c) The total hardness and yield strength in terms of Nb concentration for a fixed 87 mole \% Ti.}\label{fig:TiAlNbstrength}
\end{figure}
\begin{figure}[!h]\includegraphics[width=\textwidth]{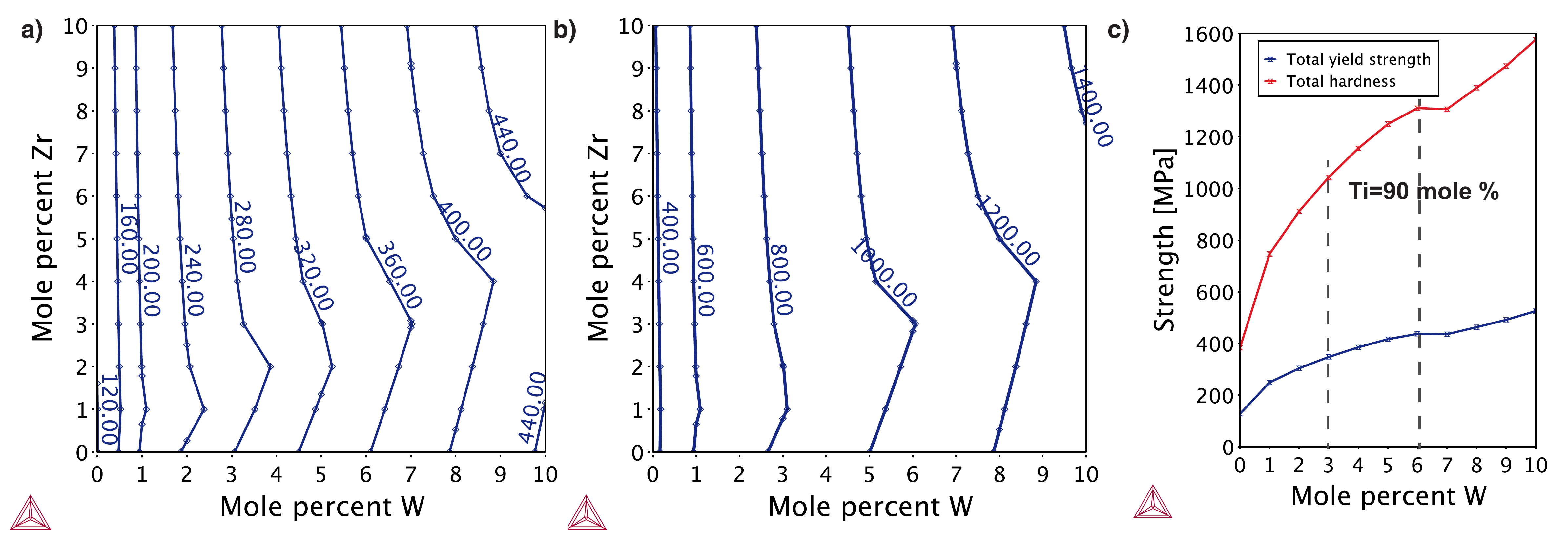}
\caption{Yield strength and hardness calculation for Ti-Zr-W. a) The total yield strength and b) the total hardness contours in terms of Zr and W concentrations. c) The total hardness and yield strength in terms of W concentration for a fixed 90 mole \% Ti.}
\label{fig:TiZrWstrength}
\end{figure}
%
Using the Property Model module of ThermoCalc, we calculated the total yield strength and the total hardness of the selected alloys. We include the solid solution, grain boundary, and precipitate strengthening and hardening effects for calculating the total yield strength and hardness, respectively. We plot contours of yield strength and hardness in terms of alloy composition to guide us target a composition that maximizes the total hardness. 
Figures \ref{fig:TiAlVstrength} (a) and (b) illustrate the total yield strength and hardness contours, respectively, in terms of aluminum and vanadium concentrations for the Ti-Al-V ternary alloy. Note that titanium concentration is equal to $X_{Ti }= 1 - X_{Al} - X_{V}$, and as we move along the \textit{x} and \textit{y} axes, titanium concentration varies. Both the total yield strength and hardness strongly dependent on vanadium concentration, as opposed to their weak dependence on aluminum. Figure \ref{fig:TiAlVstrength} (c) illustrates the total hardness and yield strength in terms of Al concentration for a fixed mole fraction of 0.9 Ti. For the benchmark system of Ti-6Al-4V (mole \%), the calculated yield strength and hardness are 271 MPa and 814 MPa, respectively. Figures \ref{fig:TiAlNbstrength} and \ref{fig:TiZrWstrength} illustrate the yield strength and hardness calculations for Ti-Al-Nb and Ti-Zr-W, respectively. For the Ti-Al-Nb system, both yield strength and total hardness strongly depend on Nb concentration if Al concentration is above 3 mole \% (see Figure \ref{fig:TiAlNbstrength} (a-b)). For Al concentration below 3 mole \%, the situation is reversed and Nb concentration has negligible effect on yield strength and hardness. As shown in Figures \ref{fig:TiAlNbstrength}(c), Ti-Al-Nb hardness increases with Nb concentration with a progressively decreasing slope up to 6-7 mole \% Nb and then decreases with further Nb concentration. Therefore, we select a composition of 6 mole \% Nb to maximize the hardness.  For the Ti-Zr-W system, both yield strength and total hardness strongly depend on W concentration with a weak dependence on Zr concentration. As shown in Figures \ref{fig:TiZrWstrength}(c), TiZrW hardness and yield strength monotonically increase with W concentration. We select a composition of Ti-7Zr-3W (mole \%) with a yield strength and hardness of 347.8 MPa and 1043.4 MPa, respectively. Similar plots of yield strength and hardness for other alloys of Table~\ref{mytable:table1} are shown in Supplemental Figures 6 to 10. The yield strength and hardness values with the corresponding contribution from precipitate phases are summarized in Table~\ref{mytable:table2}. The yield strength and hardness of typical titanium implant bars of different ASTM grades are also shown in Table~\ref{mytable:table2} for comparison (values are extracted from references~\cite{Elias2008,Sidambe2014}). 

\begin{mytable}[!htp]\includegraphics[width=\textwidth]{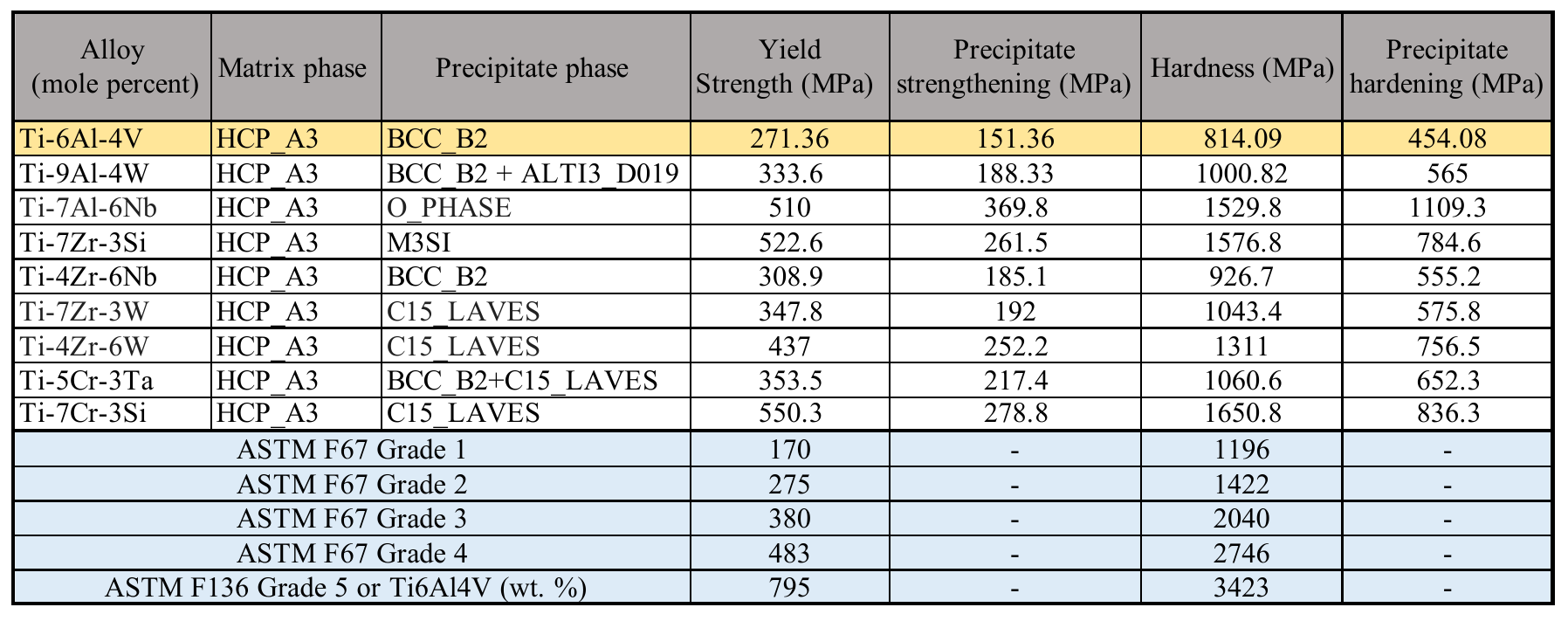}
\caption{The calculated values of yield strength and hardness for selected titanium alloys in this study compared with typical titanium implant bars. The precipitate strengthening and hardening are also reported. The blue-colored rows are values for different ASTM grades of titanium alloy. The reported values for  ASTM grades 1 to 5 are from the American Metal Society (ASM) \textit{matweb} database~\cite{Elias2008,Sidambe2014}. Hardness values are converted from Vickers to MPa. }
\label{mytable:table2}
\end{mytable}

All the selected alloys in this study have a relatively yield strength and an increased hardness compared to the benchmark system of Ti-6Al-4V (mole \%). The goal of the comparison of the selected alloys with the benchmark system is to understand the effect of the second and third alloying elements on changing the yield strength and hardness. The improved hardness in the selected alloys is compatible with our predictions based on atomic and electronic parameters in tier one. For example, by including dissolution blocker elements, such as Nb, Ta, and W (see Figure~\ref{fig:DFT1}), or elements with high values of Teter hardness, such as Cr, Mo, and W (see Figure~\ref{fig:DFT2}), as the third alloying elements instead of V, the total hardness has increased compared to Ti-6Al-4V. A significant contribution that is not present in atomic level predictions is the precipitate hardening or strengthening, which arises due to a precipitate phase in the alloy. As shown in Table~\ref{mytable:table2}, the precipitate hardening or strengthening accounts for about 50-60 \% of the total hardness or strength, except for Ti-7Al-6Nb, where the precipitate contribution is about 72\%. This is due to the existence of a highly ordered orthorhombic phase (O PHASE) as the precipitate. 
%

\subsection{Protective Oxide Scale formation}\label{sec:pourbaix}
To assess the corrosion resistance of the selected alloys, we adapt a simple metric introduced by Pilling and Bedworth, who divided oxidizable metals into two groups: those that form protective oxide scales and those that do not~\cite{Bradford1993}. They defined the Pilling-Bedworth ratio as the volume of the oxide divided by the volume of the metal that formed the oxide. They proposed that if the ratio is less than 1, the oxide scale is not protective because the volume of the oxide is insufficient to cover the entire underlying metal surface. Typical examples of metals that form oxides with PB ratio lower than 1 are alkali and alkaline earth metals \cite{Bradford1993}. If the ratio is greater than 1 and lower than 2, the oxide scale is protective as it shields the entire metal surface with a low risk of cracking due to volume mismatch. Aluminum is a good example of a metal that forms a protective oxide ($\alpha$\ce{Al2O3}), with a PB ratio of 1.28~\cite{Bradford1993}. A PB ratio much greater than 2 leads to large compressive stress built up in the oxide and consequent buckling and spalling off, rending the oxide an unprotective scale. Iron is a good example of a metal that forms an unprotective oxide ($\alpha$\ce{Fe2O3}), with a PB ratio of 2.15~\cite{Bradford1993}. 

As we move from pure metal systems to alloys, the Pilling-Bedworth ratio calculation becomes more complicated. For alloys, both the underling metallic system and the formed oxide may have a multi-phase nature. Additionally, the oxidation reaction differs by varying the thermodynamic (temperature or pressure) and the electro-chemical (acidity or electric potential) conditions of the environment. For example, as an alloy is exposed to the aqueous environment of the body, it undergoes an oxidation reaction of the following typical form

\begin{align*}
\ce{{M}_{1-x-y} {M'}_{x}{M''}_{y} (s) + 2 H2O (l) -> & (1-x-y) {M}O2 +x {M'}O2 +y {M''}O2 \\ & + 4H^+ + 4e^-}
\end{align*}

where $M$, $M'$, and $M''$ represent three different metallic elements in a ternary alloy. \red{Here, we consider the aqueous environment to be pure water. We assume the effects of other chemical components, such as NaCl and HCl in human blood solution, on the products of the oxidation reaction is negligible.} The production of hydrogen ions and free electrons make the aqueous oxidation an electrochemical reaction which can in turn change the stability of the formed oxides. The formed oxides can have multiple phases or other phases of oxides such as \ce{{MM'}O2} can form.  
\begin{figure}[!h]\includegraphics[width=\textwidth]{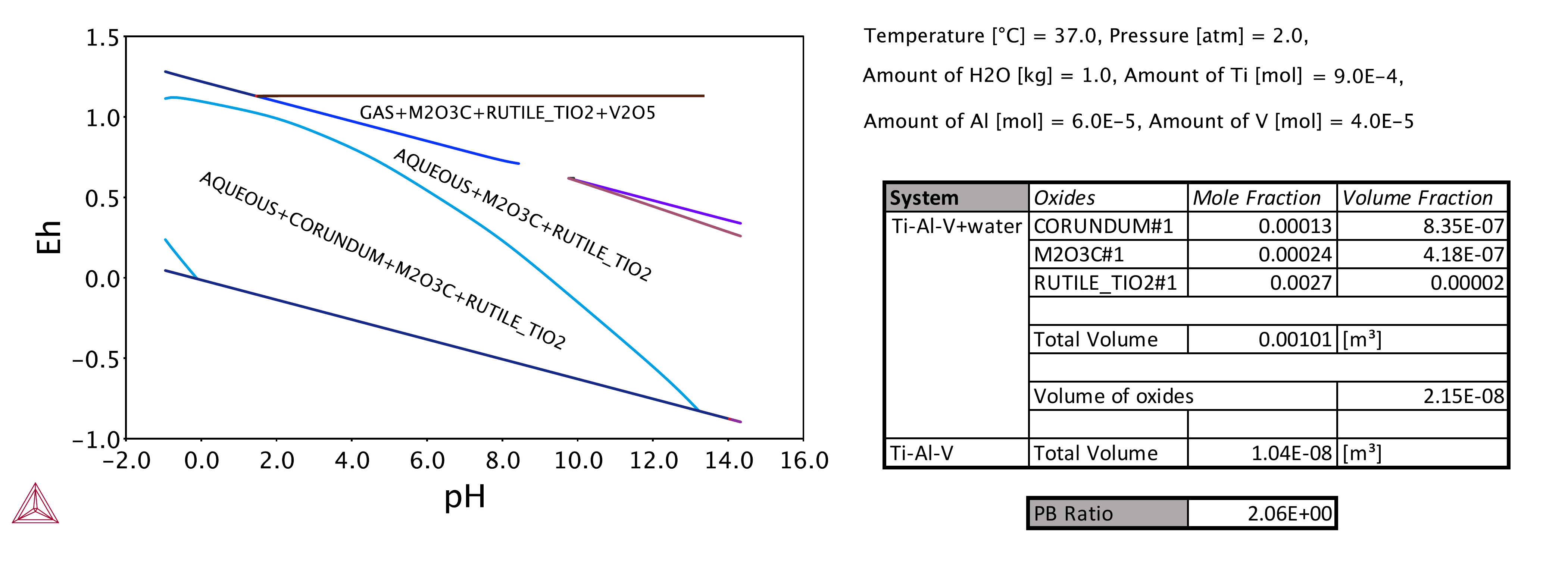}
\caption{Protective oxide scale formation assessment for the Ti-Al-V alloy. a) The Pourbaix diagram and b) the PB ratio for 90Ti-6Al-4V (mole \%). }
\label{fig:Pourbaix-PB-TiAlV}
\end{figure}
To overcome these complications, we use the Pourbaix diagram to predict the formed oxide in the aqueous environment of the body. The Pourbaix diagram represents the thermodynamically stable oxide phases formed in an aqueous solution in terms of the metal-solution electric potential (Eh) and the acidity of the solution (pH). We utilize ThermoCalc to calculate the Pourbaix diagram for the selected alloys at 37$^{\circ}$C, 2 atm, and $10^{-3}$ mole of the underlying alloy in the presence of 1 Kg of water. Figure~\ref{fig:Pourbaix-PB-TiAlV} (a) illustrates the Pourbaix diagram for Ti-6Al-4V (mole \%) as an example.  We assume that the pH and Eh for the body environment are 7.4 and 0.15 volt, respectively~\cite{VIRTANEN2018128,Abramson1936,Bondar2012}. \red{The pH value used here accounts for the acidity of human blood solution containig NaCl and HCl}. We then identify the stable oxides from the Pourbaix diagram at the designated pH and Eh. 

\begin{figure}[!h]\includegraphics[width=\textwidth]{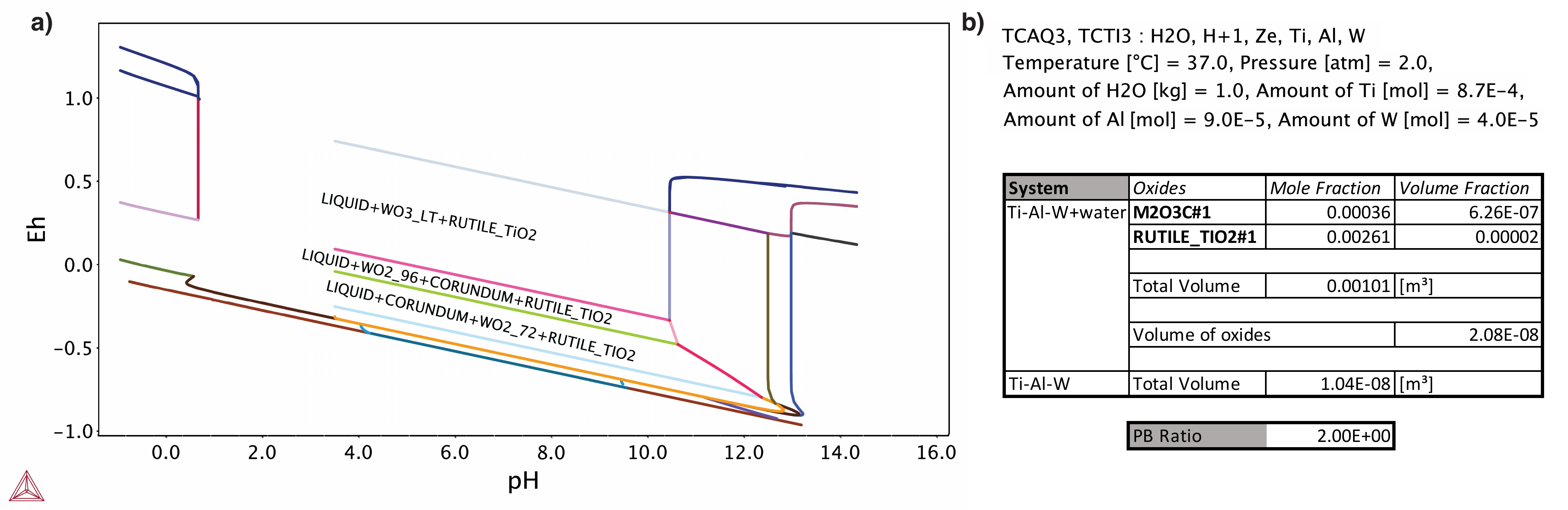}
\caption{Protective oxide scale formation assessment for the Ti-Al-W alloy. a) The Pourbaix diagram and b) the PB ratio for 87Ti-9Al-4W (mole \%). }
\label{fig:Pourbaix-PB-Ti-Al-W}
\end{figure}

\begin{figure}[!h]\includegraphics[width=\textwidth]{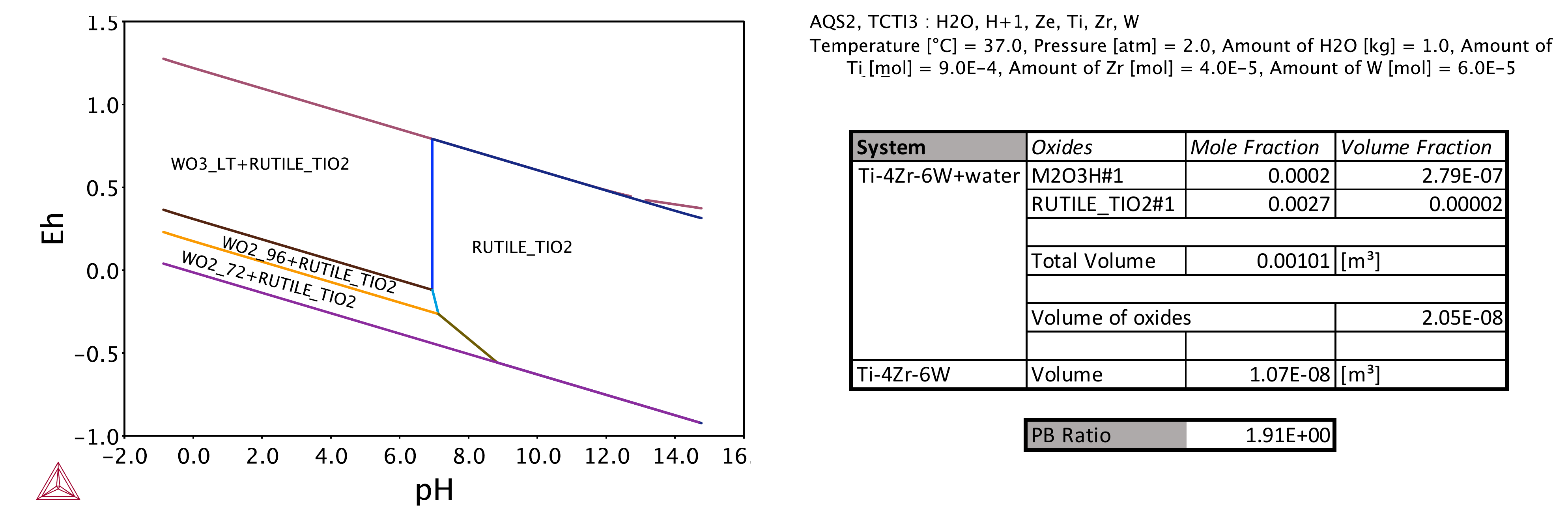}
\caption{Protective oxide scale formation assessment for the Ti-Zr-W alloy. a) The Pourbaix diagram and b) the PB ratio for 90Ti-4Zr-6W (mole \%). }
\label{fig:Pourbaix-PB-Ti-Zr-W}
\end{figure}

In order to calculated the PB ratio for the selected alloys, we first identify the stable oxides from the Pourbaix diagram at the estimated pH and Eh values of the human body. We then calculate the volume of the formed oxide phases and the volume of the underlying alloy phases to determine the PB ratio. Figure~\ref{fig:Pourbaix-PB-TiAlV} (b) illustrates the calculated volume of the formed oxides and the underlying alloy phases and the PB ratio for Ti-6Al-4V (mole \%). Figures \ref{fig:Pourbaix-PB-Ti-Al-W} and \ref{fig:Pourbaix-PB-Ti-Zr-W} show the Pourbaix diagram and the PB ratio calculation for Ti-9Al-4W (mole \%) and Ti-4Zr-6W (mole \%), respectively.

We could not calculate the Pourbaix diagram for all of the selected alloys because of a lack of data. However, we could calculate the PB ratios of all the selected alloys. The Pourbaix diagrams and PB ratios are reported in Supplemental Figures 11 and 12 and Supplemental Table 2. 

The calculated PB ratio, yield strength and hardness for all the selected alloys are summarized in Table~\ref{mytable:table3}. All the selected alloys in this study have a PB ratio lower than 2 as for the benchmark system Ti-6Al-4V (mole \%) and therefore are more resistant to corrosion. The PB ratio, yield strength and hardness are the three design parameters we use simultaneously to assess the corrosion and wear resistance of the alloys. As shown in the table, we ranked the alloys from the best to the least desired based on their performance as an implant. The best alloy, Ti-3Si-7Zr, shows a PB ratio of 1.49, which is an ideal value between 1 to 2 showing that the formed oxide scale is protective. Additionally, the selected Ti-Si-Zr alloy maximizes the hardness for maximum wear resistance. 

\begin{mytable}[!htp]\includegraphics[width=\textwidth]{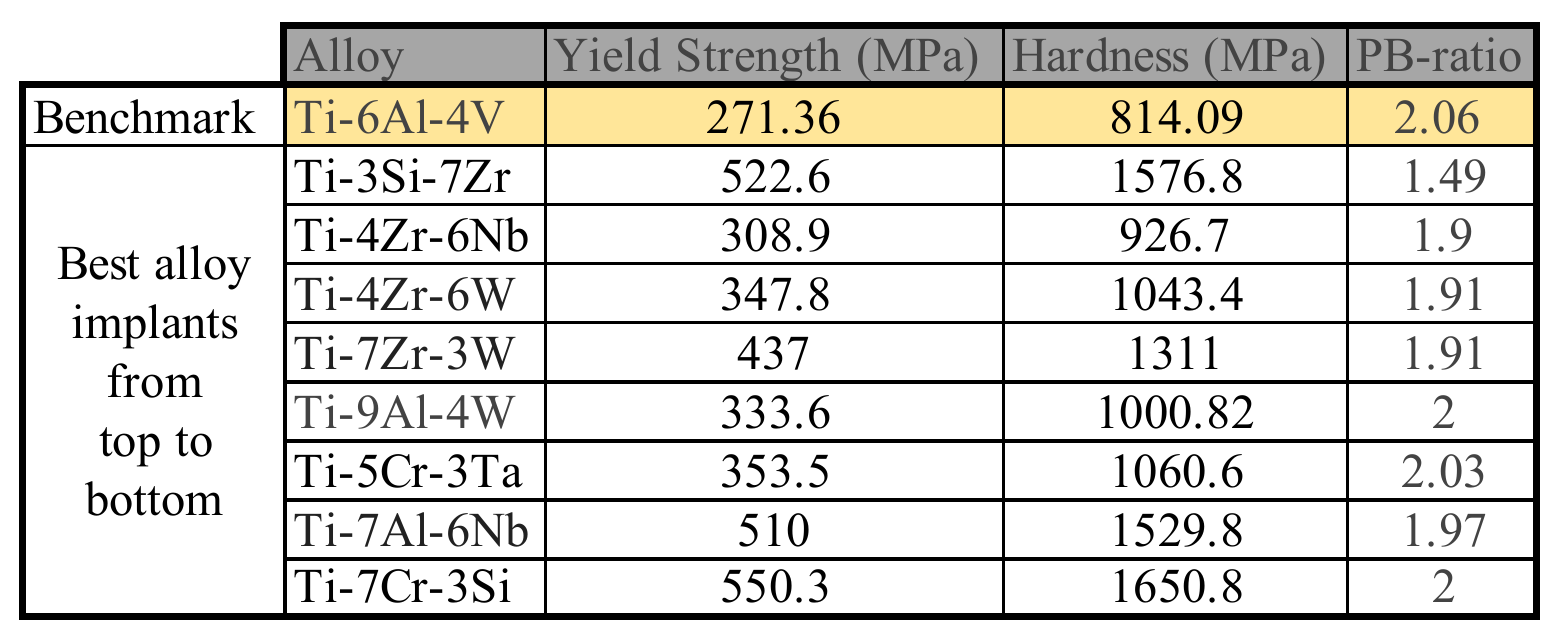}
\caption{The yield strength, hardness, and Pilling-Bedworth (PB) ratio as the design parameters for improving the wear and corrosion resistance of titanium alloys.}
\label{mytable:table3}
\end{mytable}

\section{Towards Silicon Alloys}\label{sec:silicon}
We further investigate the best selected alloy in our study, the Ti-Si-Zr ternary system as shown in Table \ref{mytable:table3}, to understand the role of silicon in enhancing the wear and corrosion resistance of the alloy. According to the atomic parameters shown in Figure \ref{fig:DFT1} (b), silicon is an oxide scale promoter due to its high oxide formation tendency and has the highest surface work function among our selected alloying elements and therefore is resistant to galvanic corrosion. Additionally, based on the prediction of the Teter hardness in Figure \ref{fig:DFT2} (b), pure silicon is relatively hard. These desirable properties make Si a good alloying candidate. As we show in the following, one advantage of alloying the system with silicon is the significant improvement of the corrosion resistance. This is because Si is immiscible with elements such as Ti and Zr and the alloy forms many intermediate compounds and has a multi-phase nature. Additionally, Si is a metalloid and forms highly directional covalent bonds while avoiding to form close packed structure unlike a metallic alloy. Therefore, the underlying crystal structure is relatively less dense with a large unit cell volume. Therefore, the volume of the underling Si alloy and the formed oxides are closer than of a close packed alloy and the formed oxides. This volume similarity can significantly enhance corrosion resistance. 
 
 \begin{figure}[h]\includegraphics[width=\textwidth]{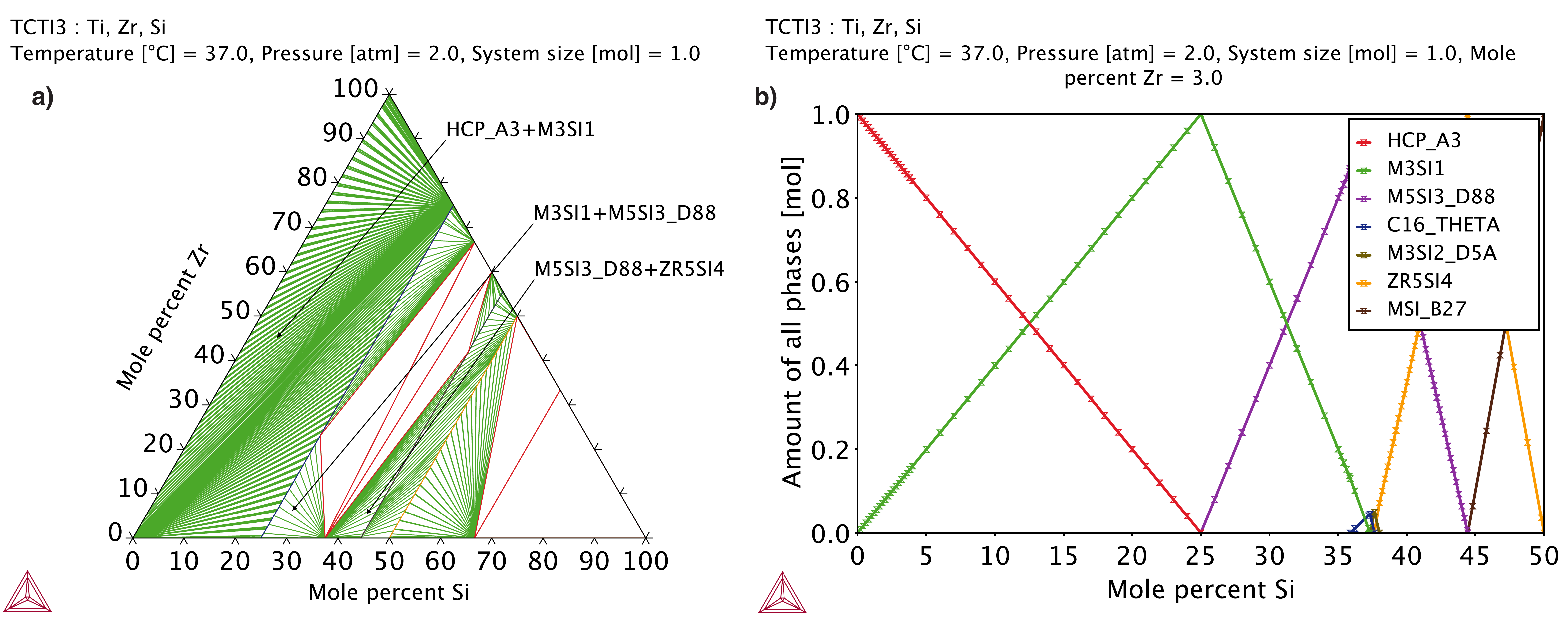}
\caption{Phase stability assessment for the Ti-Si-Zr alloy. a) The ternary phase diagram at 37$^{\circ}$C and 2 atm, with tie lines shown by green and phase boundaries shown by red. b) The mole fraction of stable phases in terms of Si concentration for a fixed 3 mole \% Zr. }
\label{fig:TiZrSi-stability-detail}
\end{figure}
Figure \ref{fig:TiZrSi-stability-detail} (a) shows the ternary phase diagram of Ti-Si-Zr, which confirms the formation of intermediate compounds and the multi-phase nature of the alloy. Figure \ref{fig:TiZrSi-stability-detail} (b) illustrates the mole fraction of stable phases as a function of silicon concentration with Zr fixed at 3 mole \%. As shown, Ti-Si-Zr system forms several intermediate compounds such as M3SI (cubic bismuth trifluoride structure), M5SI3 (hexagonal $P6_3/mcm$ structure), and ZR5SI4 (Zirconium(IV) silicate). We expect that the formed compounds enhance the hardness and strength due to precipitate hardening and strengthening effects.  

\begin{figure}[h]\includegraphics[width=\textwidth]{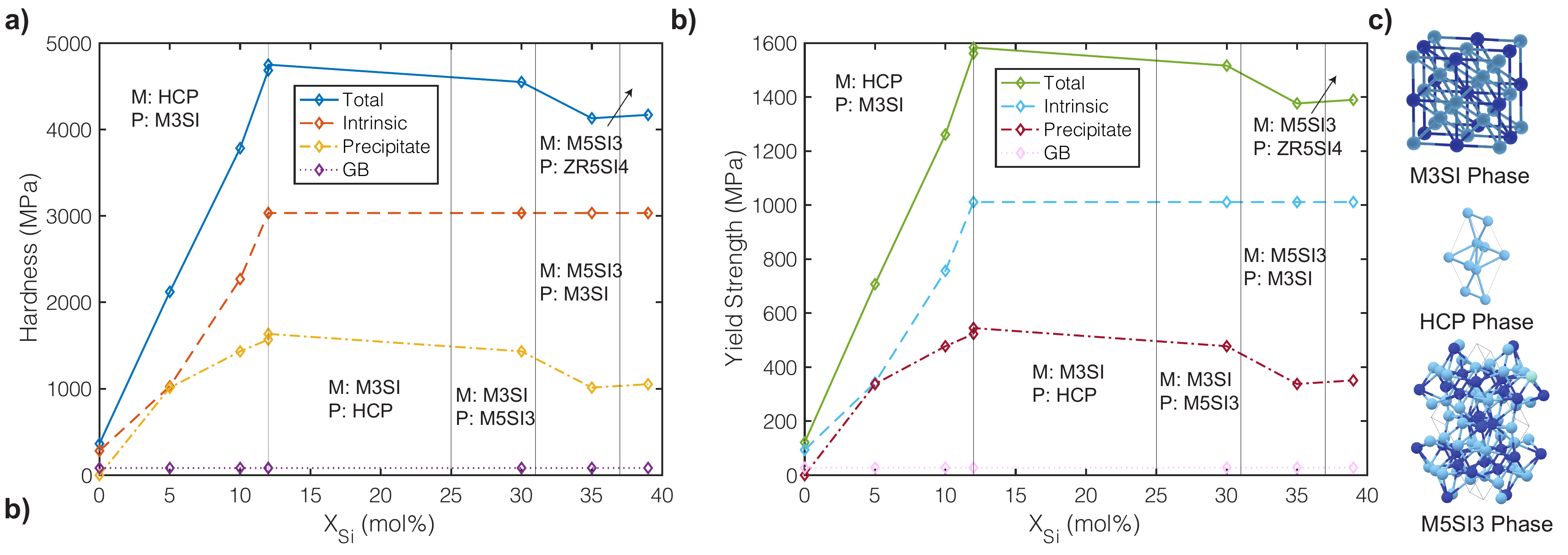}
\caption{Strength of Ti-Si-Zr alloy versus silicon concentration. a) Total hardness and b) yield strength of Ti-Si-Zr alloy at a fixed concentration of 3 mole.\% Zr in terms of silicon concentration. The vertical lines divide different two-phase Ti-Si-Zr alloys with different matrix and precipitate phases. c) The crystal structure of the relevant phases for the Ti-Si-Zr system. M3SI has a bismuth trifluoride structure, M5SI3 has hexagonal $P6_3/mcm$ structure, and HCP is the hexagonal close packed structure.}
\label{fig:Ti-Si-Zr-Hardness-Yiled}
\end{figure}
Figure \ref{fig:Ti-Si-Zr-Hardness-Yiled} illustrate the variation of hardness and yield strength as a function of Si concentration and confirms that addition of Si increases the hardness and yield strength of the alloy. Starting from zero silicon concentration, the total hardness and yield strength increases by alloying Si. As shown in Figure \ref{fig:Ti-Si-Zr-Hardness-Yiled}, the increase in hardness (and yield strength) by adding Si up to 12 mole\% stems from an increase in the intrinsic hardness (and strength) as well as an increase in the precipitate hardening (and strengthening). The intrinsic hardness (and strength) of the M3SI phase is larger than HCP (approximated by $H=0.151G$ of 11 GPa for \ce{Ti3Si} and 9.6 GPa for hcp Ti where $G$ is extracted from the AFLOW-ML prediction~\cite{Isayev2017}). Addition of Si results in the progressive formation of M3SI precipitates (as shown in Figure \ref{fig:TiZrSi-stability-detail} (b)) increasing both the intrinsic hardness (and strength) and precipitate hardening (and strengthening). The TCTI3 database in ThermoCalc lacks the intrinsic hardness and strength data for the M3SI and M5SI3 phases. Therefore, we replicate the intrinsic hardness and yield strength of 12 mole\% Si for other concentrations up to 40 mole.\%. As shown in \ref{fig:Ti-Si-Zr-Hardness-Yiled}, the precipitate hardening and strengthening effects remain constant, around 1000 MPa and 400 MPa, respectively, for Si concentration above 12 and below 40 mole.\%. 

\begin{mytable}[h]\includegraphics[width=\textwidth]{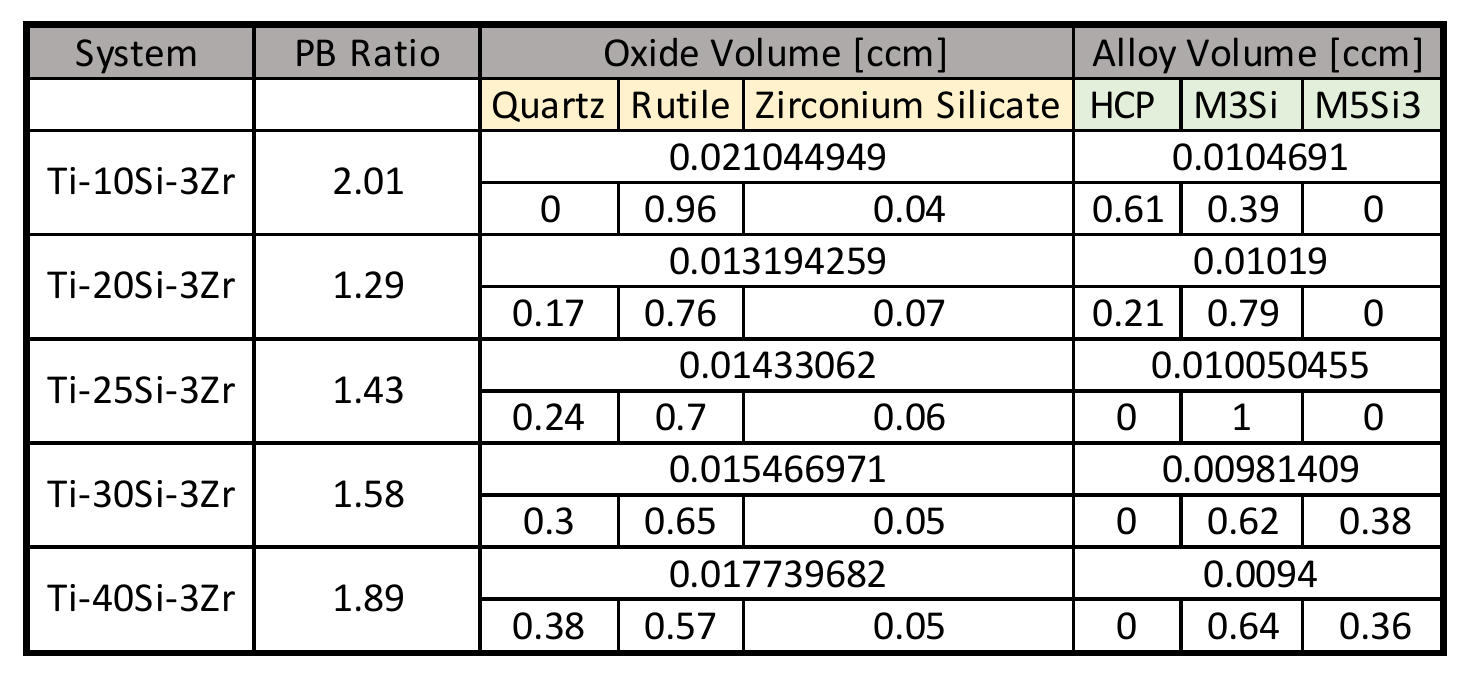}
\caption{The Pilling-Bedworth (PB) ratio for Ti-Si-Zr alloys with increased Si concentration and a fixed 3 mole.\% Zr. The volume of the oxide scale (formed from 0.001 mole of the alloy in the presence of 1 Kg water) and the underlying alloy (0.001 mole) are shown. The volume fraction of different phases of the oxide and alloy are illustrated.}
\label{mytable:table4}
\end{mytable}
To asses the corrosion resistance, we calculate the PB ratio of the Ti-Si-Zr alloy at a fixed concentration of 3 mole\% Zr and by progressively increasing Si concentration from 10 to 40 mole.\%. As shown in Table \ref{mytable:table4}, by increasing Si concentration above 10 mole.\%, quartz forms at the expense of rutile, which lead to an overall decrease of the oxide volume. The best PB ratio lies below 2 and above 1. Therefore, Si concentrations between 20 to 40 mole.\% will form protective oxide scale with enhanced corrosion resistance.

\section{Discussion}\label{sec:discussion}
\red{
The static corrosion analysis in this work is only a qualitative measure to assess the overall corrosion resistance of the alloy. This analysis aims to provide a simple metric for comparison of a large number of alloys and to down-select the ones that form passive protective oxide scales. The \textit{in vivo} corrosion behavior of metallic materials used in biomedical applications is complex and predicting the long-term corrosion requires sophisticated kinetic simulations of the corrosion process in the presence of biological conditions of the human body (e.g., body fluids, proteins, cells) and functional activities (e.g., static and dynamic loads)~\cite{Virtanen2018}. These kinetic studies are beyond the scope and purpose of this work. In this work, we only study the uniform corrosion (as compared to pitting corrosion or crevice corrosion) of the metallic materials in the water environment under static loading.}
\par
\red{Additionally, in this work, the resistance of the alloy to mechanical wear degradation and chemical corrosion are measured by two separate metrics of hardness and the Pilling-Bedworth (PB) ratio, respectively. These metrics aim to provide a simple and high-throughput method to assess the combination of wear and corrosion resistance. A more realistic analysis is to study the tribocorrosion degradation by combining the corrosion and wear degradation processes~\cite{Yan2006}. However, tribocorrosion experimental analysis in simulated human body conditions~\cite{WELLES2021e07023,SIMOES2016410,Cheng2019,BRONCZYK2019202966,YAN20071105,MARQUES20168} or numerical simulations of tribocorrosion~\cite{FALLAHNEZHAD2022107284} are time-consuming and cannot serve as a high-throughput measure for down-selecting implant materials. Another avenue that our study does not investigate in depth is the bio-compatibility of our output alloys. We assume a high level of bio-compatibility in our alloys as similar materials in literature have been tested and have produced data agreeing with our assumptions \cite{OZAN2017119}.
}

\red{
The results of the our analysis are consistence with the experimental observations of the corrosion behavior and wear resistance of titanium alloys for orthopedic implants. A recent comparative review among 189 articles showed that the addition of niobium, tantalum and zirconium as alloying elements to beta titanium alloys increases their corrosion protection~\cite{Juliana2020}. Additionally, aiming to stabilize the beta phase through the use of a stabilizing element such as niobium creates an alloy with higher strength and ductility \cite{CHUI201754}. Zirconium in particular, has been shown to increase hardness when alloyed with titanium \cite{Ho2008}. Laser-deposited alloys of Ti-35Nb-7Zr-5Ta (wt \%) exhibit excellent corrosion resistance compared to CP Ti (Grade 2) alloys, as shown by \textit{In vitro} corrosion studies~\cite{Samuel2010}. These observations are consistent with our tier-1 approach of selecting dissolution blockers (i.e., elements with high cohesive energies as shown in Fig.~\ref{fig:DFT1} (b)) as alloying elements to improve the wear and corrosion resistance (see the results in Tables~\ref{mytable:table2} and \ref{mytable:table3}). One approach to improving the stability of the resultant alloy that this study does not look at is the final finish of the material as well as manufacturing techniques to obtain the best result. For corrosion resistance, we only consider the oxide scale formed by the alloy, however methods such as ceramic coating and/or mechanical polishing have been shown to work in conjunction with the oxide scale to further improve corrosion resistance \cite{KajzerAEvaluation2016,Li2004kz,Xin2021}. More advanced manufacturing practices such as a sputtering process can alter the oxide scale to obtain a more favorable phase ratio \cite{FRUTOS201844}. The outcomes of our high-throughput method can be a basis for new research in this field. 
}

\section{Conclusion}\label{sec:conclusion}
We introduce a high-throughput method based on simple parameters of pure elements, namely the cohesive energy, oxide formation energy, surface work function, and elastic shear modulus, to select alloying elements that improve the wear-resistance and corrosion-resistance of alloys. We illustrate that predictions based on theses simple parameters agree well with a detailed analysis of wear and corrosion resistance based on a comprehensive CALPHAD approach, thereby providing useful metrics for rapid screening of alloys for improved corrosion and wear resistance. 
%
In the detailed analysis, we assess the wear and corrosion resistance of the alloy based on hardness and Pilling-Bedworth (PB) ratio measures, respectively. We calculated these measures based on a comprehensive thermodynamic analysis of the underlying alloy phases and the formed oxides in the scale using the ThermoCalc software, which integrates the CALPHAD database with optimization tools.  

In general, alloying elements with a high intrinsic hardness predicted from DFT, such as Cr, Ta, and W, and those forming highly ordered precipitate, such as Si and Nb, can significantly increase the hardness and thereby the wear resistance of the alloy. \red{Additionally, we optimized the composition of the alloys in this study by a synergetic combination of alloying elements that are oxide scale promoters and dissolution blockers. Accordingly,} all the modified alloys compared to the Ti-6Al-4V (mole \%) benchmark show a lower PB ratio, indicating their improved corrosion-resistance. Additionally, we showed that Si as an alloying element, especially at close to equal concentrations to Ti, can significantly reduce the PB ratio and therefore improve the corrosion resistance and wear resistance of the alloy. This observation opens the door towards using Si alloys for orthopedic implants. 
\section{Acknowledgement}
Sara Kadkhodaei acknowledges the National Science Foundation (NSF) Award Number DMR-1954621 for financial support. Additionally, this work was supported by the US Department of Defense, National Defense Education Program through the Educational and Research Training Collaborative at the University of Illinois Chicago, Award Number HQ00342010037. The ThermoCalc software and databases used in this work are part of the Materials Genome Toolkit awarded to UIC by the American Society for Metals (ASM) Materials Education Foundation. 

\section{Data Availability}
The raw data required to reproduce these findings, namely the TCTI3 database for thermodynamic and atomic mobility data, cannot be shared and are only available on ThermoCalc to licensed users.

\section{Author Contribution}
Noel Siony: Methodology, Investigation, Writing - Original draft preparation, Visualization. Long Vuong: Methodology, Investigation, Writing - Original draft preparation, Visualization. Otgonsuren Lundaajamts: Methodology, Investigation, Writing - Original draft preparation, Visualization. Sara Kadkhodaei: Conceptualization, Methodology, Investigation, Writing - Review and editing, Supervision, Funding acquisition. 


\end{document}